\documentclass[aps,prl,superscriptaddress,twocolumn,floatfix,showpacs,longbibliography,nobibnotes]{revtex4-1}

\usepackage{graphicx,color}
\usepackage[utf8]{inputenc}
\usepackage{amsmath,amssymb,bm,amsfonts,dsfont,mathrsfs,amsthm}
\usepackage{gensymb}
\usepackage{braket}
\usepackage{txfonts,comment}
\usepackage[colorlinks=true,linkcolor=blue,citecolor=blue,urlcolor=blue]{hyperref}
\usepackage{soul}
\usepackage{enumitem}
\usepackage{orcidlink}

\newcommand{\nn}{\nonumber \\}

\AtBeginDocument{%
    \newwrite\bibnotes
    \def\bibnotesext{Notes.bib}
    \immediate\openout\bibnotes=\jobname\bibnotesext
    \immediate\write\bibnotes{@CONTROL{REVTEX41Control}}
    \immediate\write\bibnotes{@CONTROL{%
    apsrev41Control,author="08",editor="1",pages="1",title="0",year="1"}}
     \if@filesw
     \immediate\write\@auxout{\string\citation{apsrev41Control}}%
    \fi
}%

\newcommand{\beginsupplement}
    {
    \clearpage
    \onecolumngrid
    \renewcommand{\thesection}{\Alph{section}}
    \renewcommand{\thefigure}{S\arabic{figure}}
    \renewcommand{\thetable}{S\Roman{table}}
    \setcounter{figure}{0}
    \renewcommand{\theequation}{S\arabic{equation}}
    \setcounter{equation}{0}
    }

\begin{document}

\title{Observation of Criticality-Enhanced Quantum Sensing in Nonunitary Quantum Walks}

\author{Lei Xiao}
\thanks{These authors contributed equally to this work.}
%\email{xiaoleiphys@seu.edu.cn}
\affiliation{School of Physics, Southeast University, Nanjing 211189, China}
\affiliation{Key Laboratory of Quantum Materials and Devices of Ministry of Education, Southeast University, Nanjing 211189, China}

\author{Saubhik Sarkar\,\orcidlink{0000-0002-2933-2792}}
\thanks{These authors contributed equally to this work.}
%\email{saubhik.sarkar@uestc.edu.cn}
\affiliation{Institute of Fundamental and Frontier Sciences, University of Electronic Science and Technology of China, Chengdu 611731, China}
\affiliation{Key Laboratory of Quantum Physics and Photonic Quantum Information, Ministry of Education, University of Electronic Science and Technology of China, Chengdu 611731, China}

\author{Kunkun Wang}
%\email{kunkunwang@126.com}
\affiliation{School of Physics and Optoelectronics Engineering, Anhui University, Hefei 230601, China}

\author{Abolfazl Bayat\,\orcidlink{0000-0003-3852-4558}}
\email{abolfazl.bayat@uestc.edu.cn}
\affiliation{Institute of Fundamental and Frontier Sciences, University of Electronic Science and Technology of China, Chengdu 611731, China}
\affiliation{Key Laboratory of Quantum Physics and Photonic Quantum Information, Ministry of Education, University of Electronic Science and Technology of China, Chengdu 611731, China}
\affiliation{Shimmer Center, Tianfu Jiangxi Laboratory, Chengdu 641419, China}

\author{Peng Xue\,\orcidlink{0000-0002-4272-2883}}
\email{gnep.eux@gmail.com}
\thanks{Present address: Beijing Computational Science Research Center, Beijing 100193, China.}
\affiliation{School of Physics, Southeast University, Nanjing 211189, China}

% =============================================================================
\begin{abstract}

Quantum physics enables parameter estimation with precisions beyond the capability of classical sensors. Quantum criticality is a key resource for this quantum-enhanced sensing, but experimental realization has been challenging due to the complexity of ground-state preparation and the long time required to reach the steady state near criticality. Here, we experimentally demonstrate critical enhancement in a non-Hermitian topological system using a photonic quantum walk setup. Our system supports two distinct phase transitions at which enhanced sensitivity is observed even at transient times for which the Bayesian inference shows excellent estimation and precision. It is a direct demonstration of criticality-enhanced scaling laws with non-unitary dynamics. 

\end{abstract}

\maketitle

% \tableofcontents

% =============================================================================
\emph{Introduction.---}
Quantum sensing harnesses quantum features to outperform classical sensing procedures~\cite{degen2017quantum}. 
The precision of estimating an unknown parameter, quantified by variance, in general scales as $N^{-b}$, where $N$ is the probe size. 
In the absence of quantum resources, such as entanglement or squeezing, the best achievable precision is limited to $b{=}1$, also known as the standard quantum limit~\cite{caves1981quantum}. 
Exploiting quantum features can lead to $b{>}1$, known as quantum-enhanced sensing ($b {=} 2$ is popularly known as the Heisenberg limit)~\cite{giovannetti2004quantum, giovannetti2011advances, deng2024quantum, chen2024quantum}.  
One distinguished resource for enhancement is quantum criticality~\cite{zanardi2008quantum, rams2018limits, wu2024experimental}, which exploits the extreme parameter-susceptibility near a phase transition in both closed or open systems~\cite{montenegro2025review}. 
However, practical implementation is severely limited by critical slowing down during ground state preparation~\cite{sachdev1999quantum} and long evolution time to reach steady states~\cite{macieszczak2016towards, garbe2020critical}.
Therefore, it is highly desirable to observe definitive signatures of critical enhancement within an experimentally accessible timescale.

So far, criticality-enhanced quantum sensing has been experimentally implemented for: (i) first-order phase transitions in nuclear magnetic resonance systems~\cite{liu2021experimental} and superconducting circuits~\cite{yu2025experimental}; (ii) second-order phase transitions with Rydberg atoms~\cite{ding2022enhanced} and superconducting resonators~\cite{beaulieu2025criticality}.
While the system size is kept fixed in Refs.~\cite{liu2021experimental, ding2022enhanced, yu2025experimental}, Ref.~\cite{beaulieu2025criticality} reports enhanced growth of precision with rescaled Hamiltonian parameters.
For a direct demonstration of the super-linear scaling, the potential of topological phase transitions has been reported~\cite{Sarkar2022Free}. 
In particular, non-Hermitian (NH) topological models have attracted a lot of interest in recent times~\cite{Ashida2020Non, Bergholtz2021Exceptional, Kawabata2019Symmetry, Lee2016NonHermitian, Yao2018Edge, Dora2022pt, Okuma2023Non}. 
Criticality-enhanced sensing with NH systems requires topological phase transitions accompanied by closing of novel energy gaps such as point gap and line gap in the complex spectrum~\cite{sarkar2024critical}. 
However, experimental implementation of such sensors has remained elusive.  
While the required gap structure has been realized in cold atoms~\cite{Gou2020Tunable, Liang2022Dynamic}, photonic lattices~\cite{Weidemann2020Topological}, and coupled resonators~\cite{zhang2021observation, Wang2021Generating}, the highly controllable photonic quantum walk setup has the capability to capture the relevant topological transitions~\cite{Xiao2020Non, wang2021detecting, feis2025space}. 
This makes them a suitable candidate for realizing criticality-enhanced sensors for estimating bulk parameters of an NH Hamiltonian.
Note that, this is a direct manifestation of quantum criticality and is unrelated to the sensitivity of NH systems to boundary perturbations~\cite{Wiersig2014Enhancing, Zhang2019Quantum, budich2020non, Koch2022Quantum, McDonald2020Exponentially, Liu2016Metrology, Hodaei2017Enhanced, Lau2018Fundamental, Yu2020Experimental, Wang2020Petermann, xiao2024non, Yu2024Heisenberg}.

In this work, we report an experimental demonstration of criticality-enhanced sensing with discrete-time photonic quantum walk setup.
We realize a topological NH Hamiltonian evolution by engineering photon loss. 
At both types of topological phase transitions, we observe enhanced sensitivity, as superlinear scaling with respect to system size, for estimating the Hamiltonian parameter that drives the transitions. 
This enhancement is detected for the evolution at transient times, and Bayesian analysis of the experimental data discerns excellent estimation and near-optimal precision. 

% =============================================================================
\emph{Parameter estimation.---}
Consider a probe in quantum state $\rho_\theta$, with $\theta$ being an unknown parameter to be estimated. 
Measurement using a positive operator-valued measure (POVM) $\{\Pi_k\}$ provides the probability of obtaining the $k$-th outcome $p_k(\theta) {=} \text{Tr}[\rho_\theta \Pi_k]$, and an estimator is constructed to infer the value of $\theta$ from it.
The precision, quantified by the estimator's variance $\delta^2 \theta$ over $M$ repeated measurements, is fundamentally bounded by the Cram\'er-Rao inequality $\delta^2 \theta {\ge} 1/M F^C {\ge} 1/M F^Q$~\cite{braunstein1994statistical, paris2009quantum}.
Here the classical Fisher information (CFI) $F^C {=} \sum_k (\partial_\theta p_k)^2 \, / p_k$ determines the precision for a given measurement setup $\{\Pi_k\}$ and the quantum Fisher information (QFI) $F^Q$ is its upper bound, which is optimized over all possible POVMs.
See Supplemental Material (SM) for more details~\cite{Supp}.
\nocite{Yao2018Edge, Yokomizo2019Non, Xiao2020Non, deng2019non} % References for SM
For pure states $\rho_{\theta} {=} \ket{\psi_{\theta}} \bra{\psi_{\theta}}$, one gets $F^Q {=} 4\left(\braket{\partial_\theta \psi_{\theta}|\partial_\theta \psi_{\theta}} - |\braket{\partial_\theta \psi_{\theta}|\psi_{\theta}}|^2 \right)$~\cite{paris2009quantum}.
In the NH domain, where $\theta$ is encoded in $\rho_\theta$ by a non-unitary evolution, the probe state needs to be normalized so that measurements produce normalized probability distributions~\cite{Alipor2014Quantum, Yu2023Quantum, Xiao2020Non, Yu2024Heisenberg}.

% =============================================================================

\begin{figure}[t]
\centering
\includegraphics[width=0.95\linewidth]{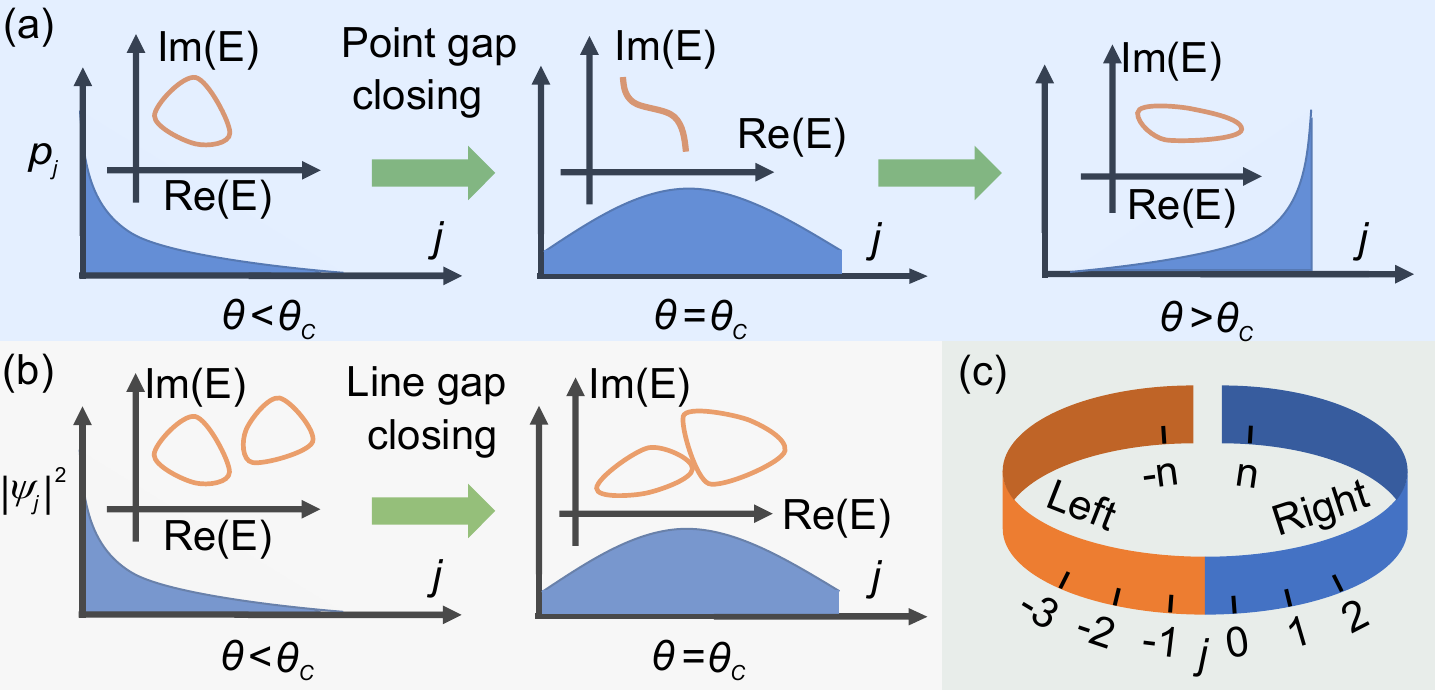} 
\caption{\textbf{Schematic of phase transition and model}. 
(a) With point gap, the complex bulk spectrum has a loop-like structure when the parameter $\theta$ is away from the critical value $\theta_c$. Correspondingly, the cumulative site population $p_j$ of all the eigenstates are edge-localized in a finite system (skin effect). At criticality, the spectral loop collapses and skin effect vanishes. 
(b) Presence of line gap supports topologically non-trivial edge states $\ket{\psi}$, localized at the edge of a finite system. At line gap closing, the parts of the spectrum join and the edge state shifts to a delocalized bulk state. 
(c) The domain wall system used in our experiment for the non-unitary photonic quantum walk. The left and right regions have different coin state rotations.}
\label{fig:scheme}
\end{figure}

\begin{figure*}[t]
\centering
\includegraphics[width=\textwidth]{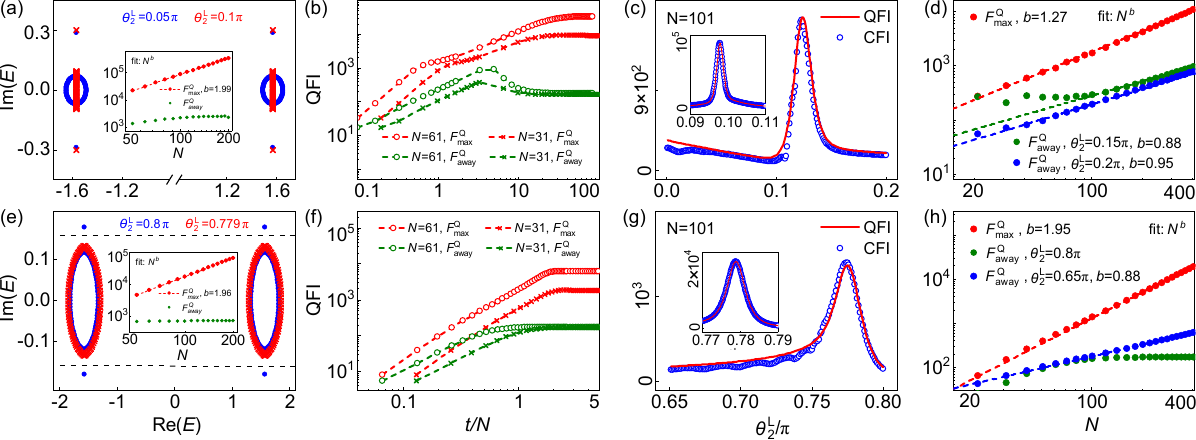} 
\caption{\textbf{Theoretical analysis}. 
(a-d) Sensing at point gap closing. The fixed parameters are $\gamma {=} 0.3$ and $\theta_1^L {=} \theta_2^R {=} 0.9 \pi$.
(a) Point gap shown for $\theta_2^L {=} 0.05 \pi$ closes at the critical value $\theta_2^L {=} 0.1 \pi$. (inset) The quadratic scaling of maximum QFI of the steady state $F^Q_{\rm max}$ with system size $N$ and absence of scaling away from criticality at $\theta_2^L {=} 0.15 \pi$. 
(b) QFI after $t$ steps of the quantum walker, at and away from criticality.
(c) QFI and CFI of $\ket{\Psi_N}$ obtained after $(N{-}1)/2$ time-steps. (inset) QFI and CFI of the steady state.
(d) Scaling of QFI of $\ket{\Psi_N}$ with $N$ (numerical fit $N^b$) near criticality and away from it ($\theta_2^L {=} 0.15 \pi, \theta_2^L {=} 0.2 \pi$).
(e-h) Sensing at line gap closing. The fixed parameters are $\gamma {=} 0.3$ and $\theta_1^L {=} \theta_2^R {=} 0.05 \pi$.
(e) Line gap shown for $\theta_2^L {=} 0.8 \pi$ with dotted lines closes at the critical value $\theta_2^L {=} 0.779 \pi$. (inset) The quadratic scaling of $F^Q_{\rm max}$ with $N$ and absence of scaling away from criticality at $\theta_2^L {=} 0.15 \pi$ for the steady state. 
(f) QFI evolution, at and away from criticality.
(g) QFI and CFI of $\ket{\Psi_N}$. (inset) QFI and CFI of the steady state.
(h) Scaling of QFI of $\ket{\Psi_N}$ near criticality and away from it ($\theta_2^L {=} 0.8 \pi, 0.65 \pi$). }
\label{fig:sim}
\end{figure*}

\emph{Non-Hermitian critical sensing.---}
The topological properties of NH Hamiltonians are connected to the complex energy gap structures, such as the point gap and line gap~\cite{Kawabata2019Symmetry}.
Gap closing at criticality is directly connected to enhanced sensitivity in Hermitian systems~\cite{abiuso2025fundamental}. 
In NH systems, closing of either type of gaps leads to distinct topological phase transitions.
A point gap occurs when the bulk eigen-spectrum creates a loop in the complex energy plane.
This corresponds to a non-trivial spectral topology and skin effect in finite systems, where eigenstates localize at the edges~\cite{Yao2018Edge}, see Fig.\ref{fig:scheme}(a).
The closure of a point gap occurs when the spectral loop (or spectral area in higher dimensionals) collapses into an arc at a critical point~\cite{Kawabata2019Symmetry}. 
This induces a spectral topological phase transition, accompanied by a localization-delocalization transition of all eigenstates, which suppresses the skin effect (see Fig.~\ref{fig:scheme}(a))~\cite{Borgnia2018Non, Okuma2020Topological, Zhang2020Correspondence, Zhang2022Universal}.
From the sensing perspective, if the probe state is taken to be a suitable eigenstate such as the steady state (with the largest imaginary part in eigenenergy), one can reach Heisenberg scaling for sensing the relevant parameter~\cite{sarkar2024critical}.
On the other hand, a line gap takes the form of a reference line that separates different parts of the spectrum, see Fig.\ref{fig:scheme}(b). 
At the closing of a line gap two disconnected parts of the spectrum join, see Fig.~\ref{fig:scheme}(b). 
This has a systematic connection with Hermitian symmetry-protected topological systems~\cite{Kawabata2019Symmetry} and results in the appearance of topologically non-trivial edge states. 
Quantum-enhanced sensitivity is again achieved when the probe is in the edge state near a line gap closing~\cite{sarkar2024critical}.

Our photonic setup realizes a quantum walk with two-dimensional coin states encoded by polarizations $H$ and $V$ on a bipartite 1D lattice of size $N {=} 2n {+} 1$, see Fig.~\ref{fig:scheme}(c). 
The evolution is governed by the non-unitary operator 
\begin{align}
    U = \begin{cases}
    R_{\theta_1^L/2} \, S \, R_{\theta_2^L/2} \, \Gamma & \quad\text{in the left part} \\ 
    R_{\theta_1^R/2} \, S \, R_{\theta_2^R/2} \, \Gamma & \quad\text{in the right part} .\\ 
    \end{cases}
    \label{QW_ham}
\end{align}
Here the coin operator $R_{\theta} {=} \sum_j \ket{j} \bra{j} \otimes e^{-i \, \theta \, \sigma_y}$ acts on the internal space with $j$ indexing the sites in either left or right part.
The different rotation angles in the left and right parts, namely $\theta_{1,2}^{L}$ and $\theta_{1,2}^{R}$, create an edge between the two domains.
The shift operator $S {=} \sum_j \ket{j-1} \bra{j} \otimes \ket{H} \bra{H} + \ket{j+1} \bra{j} \otimes \ket{V} \bra{V}$ shifts the walker to the left (right) by one site if the internal state is $\ket{H}$  ($\ket{V}$). 
Non-unitarity is caused by $\Gamma {=} \sum_j \ket{j} \bra{j}\otimes e^{\gamma \sigma_z}$.
Here $U$ provides a stroboscopic description of the dynamics under an effective NH Hamiltonian $H_{\rm eff}$, i.e.,~$U {=} e^{-i H_{\rm eff}}$~\cite{Mochizuki2016Explicit}.
For the homogeneous case (i.e., $\theta_{1}^{L,R} {=} \theta_1$ and $\theta_{2}^{L,R} {=} \theta_2$), the point gap closing in the eigen-spectrum of $H_{\rm eff}$ happens for $\cos{(\theta_1 \pm \theta_2)/2} {=} 0$~\cite{xiao2024observation}. 
We consider closed boundary conditions (CBC) to be the case when left and right edges ($j {=} {-}n$ and $j {=} n$) are connected, which is equivalent to periodic boundary conditions in a homogeneous system.
Without this connection, the system has open boundary conditions (OBC).
We estimate $\theta_{2}^{L} {=} \theta_{1}^{R}$ while $\gamma$, $\theta_{1}^{L} {=} \theta_{2}^{R}$ are fixed.

With this setup, we observe that a point gap closing condition with CBC is given by $\theta_{1}^{L} {+} \theta_{2}^{L} {=} \pi$, which applies to the homogeneous case as well (see SM~\cite{Supp}).
This is shown in Fig.~\ref{fig:sim}(a). 
As the point gap closes, all the eigenstates of the NH Hamiltonian $H_{\rm eff}$ with OBC go through the localization-delocalization phase transition. 
With the steady state as the probe, the quantum-enhancement is evident from the quadratic scaling of the peak QFI value $F^Q_{\rm max}$ with $N$ (Fig.~\ref{fig:sim}(a) inset).
The enhanced scaling disappears away from criticality for the QFI $F^Q_{\rm away}$.
The steady-state probe can be prepared by evolving the system from an arbitrary initial state for a sufficiently long time, which may not be experimentally feasible.
In our photonic setup, the quantum walker is initiated in $\ket{\Psi_0}{=}\ket{0}{\otimes} \ket{\psi_c}$, where the spatial part $\ket{0}$ is localized at the edge between the left and right domains, with an arbitrary coin state $\ket{\psi_c}$. 
While the walker needs at least $(N{-}1){/}2$ time steps to explore the whole system, reaching the steady state requires a much longer time and several reflections from the boundaries. 
The state of the system after $t$ steps is $\ket{\Psi(t)} {=} U^{t} \ket{\Psi_0}/ \, ||U^{t} \ket{\Psi_0}||$.
The QFI evolution is shown in Fig.~\ref{fig:sim}(b) near and away from criticality where three distinct features appear. 
First, the QFI shows different behaviours at different timescales: (i) monotonic increase during the transient time $t/N{<}1$ as the walker expands across the system; (ii) an irregular pre-equilibrium behavior for $1{<}t/N{<}10$ as the walker experiences reflections from the boundaries; (iii) eventual steady-state behavior beyond $t/N{>}10$. 
Second, the steady-state QFI shows quadratic scaling for the critical probe ($\theta_2^L {=} 0.1 \pi$) and becomes size-independent away from the criticality ($\theta_2^L {=} 0.15 \pi$). 
Third, in the transient time $t/N {<} 1$, the QFI of the critical probe takes significantly larger values and increases with a slope higher than the non-critical probe. 
This is a crucial observation that reveals the effect of criticality even before reaching the steady state (see SM~\cite{Supp}).
We use this feature to infer quantum-enhanced sensitivity in our experiment where a probe of size $N$ is prepared after $T {=} (N-1)/2$ steps as $\ket{\Psi_N} {=} U^{(N-1)/2} \ket{\Psi_0}/ \, ||U^{(N-1)/2} \ket{\Psi_0}||$. 
In Fig.~\ref{fig:sim}(c), we look at the QFI and the CFI of $\ket{\Psi_N}$ with $N{=}101$. 
The CFI is determined with position measurement operators $\{ \Pi_j {=} \ket{j}\bra{j} \otimes \mathds{1}_c\}$, where $\mathds{1}_c$ is the identity in the coin space.
The critical enhancement is clearly revealed by the peaks around the point gap closing for both QFI and CFI, while the inset shows the corresponding steady-state peaks. 
Additionally, as shown in Fig.~\ref{fig:sim}(d), the QFI peak value shows super-linear scaling as $F^Q_{\rm max}{\sim}N^b$ ($b{=}1.27$), in contrast to the absence of such scaling away from criticality.
The QFI attained away from criticality ($\theta_2^L {=} 0.15 \pi$) shows no scaling for small systems while sub-linear scaling is observed for large systems with $b{=}0.88$.
A different point away from criticality ($\theta_2^L {=} 0.2 \pi$) shows sub-linear scaling with $b{=}0.95$.

Similar qualitative behavior is observed for the line gap closing by identifying such scenario at $\theta_{1}^{L} {=} 0.05 \pi, \theta_{2}^{L} {=} 0.779 \pi$, see Fig.~\ref{fig:sim}(e).
Here the steady state also emerges as the edge state localized at the boundary between the domains. 
The peak value of the steady-state QFI $F^Q_{\rm max}$ shows Heisenberg scaling while the scaling disappears for $F^Q_{\rm away}$ at $\theta_{2}^{L} {=} 0.79 \pi$ (Fig.~\ref{fig:sim}(e) inset).
For the QFI evolution is shown in Fig.~\ref{fig:sim}(f), the steady-state behavior is reached after the transient time, beyond $t/N {\sim} 1$.  
Again, quadratic scaling is only obtainable for the critical steady state, whereas QFI in the non-critical case does not scale with $N$. 
In Fig.~\ref{fig:sim}(g), the transient time behavior for $\ket{\Psi_N}$ shows the signature of critical enhancement similar to the steady state (inset). 
Finally, Fig.~\ref{fig:sim}(h) displays the scaling of the critical probe to be close to the Heisenberg limit with exponent $b{=}1.95$. 
Away from criticality, the behavior changes drastically, as demonstrated by an eventual vanishing of scaling at $\theta_2^L {=} 0.8 \pi$ and sub-linear scaling at $\theta_2^L {=} 0.65 \pi$ with exponent $b{=}0.88$.
Therefore, we can detect both types of critical enhancements in experimentally accessible timescales. 

% =============================================================================
\begin{figure}[b]
\centering
\includegraphics[width=\linewidth]{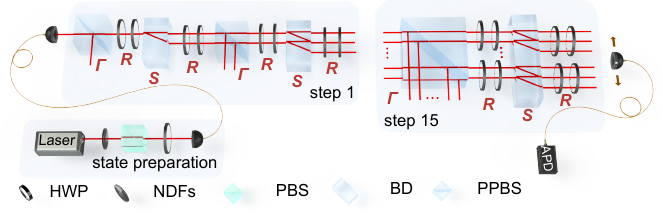}
\caption{\textbf{Experimental setup}. A pulsed laser beam is attenuated to the single-photon level. The photon polarization is then initialized in the desired state with a PBS and a HWP. It then enters a quantum walk interferometric network, where the loss operator $\Gamma$, coin operator $R$, and shift operator $S$ are implemented by the PPBS, two HWPs, and the BD, respectively. Finally, the photons are detected by APD.}
\label{fig:setup}
\end{figure}

\emph{Experimental implementation.---}
We realize the quantum walk described in Eq.~\eqref{QW_ham} using a single-photon interferometric network shown in Fig.~\ref{fig:setup}. 
Single photons, attenuated from the pulse with neutral density filters (NDFs), are prepared in the initial coin state $\ket{\psi_c}$ with polarizing beam splitter (PBS) and half-wave plate (HWP).
Non-unitarity is realized using a partially polarizing beam splitter (PPBS), which probabilistically reflects the $\ket{V}$ state and fully transmits the $\ket{H}$ state.
The position-dependent coin operator $R_{\theta}$ is implemented at each site with two HWPs set at different angles in the left and right regions.
A birefringent beam displacer (BD), which transmits horizontally (vertically) polarized photons with (without) deviation, implements the shift operator $S$. 
Final photon numbers at each site are counted with avalanche photodiodes (APDs).
Further details are presented in the \textit{End Matter}.

In our experiment, the initial state is $\ket{\Psi_0} {=} \ket{0} {\otimes} \ket{\psi_c}$, and evolves to $|\Psi_N\rangle {=} U^{(N-1)/2} |\Psi_0\rangle$ over $T {=} (N {-} 1)/2$ discrete steps. 
We characterize the normalized probability distribution of the measurements on the state $|\Psi_N\rangle$ via $\tilde{p}_j {=} \frac{m_j}{\sum_{j} m_j}$, where $m_j$ denotes the photon counts at site $j$.
In the asymptotic limit of photon counts, the probabilities $\tilde{p}_j$ approach the theoretical value $p_j$.
To evaluate the CFI $F^C(\theta_i) {=} \sum_j \frac{[\partial_{\theta_i} \tilde{p}_j]^2}{\tilde{p}_j}$, we determine $\partial_{\theta_i} \tilde{p}_j$ with discrete parameters, $\partial_{\theta_i} \tilde{p}_j {=} \frac{\tilde{p}_j(\theta_{i+1}) - \tilde{p}_j(\theta_i)}{\Delta \theta}$, where $i$ denotes the index of the parameter samples with increment $\Delta \theta {=} \theta_{i+1} {-} \theta_i$.
In the experiment, we are practically limited to $\Delta \theta {=}0.01\pi$. 
While this choice shows proper convergence of the QFI, it may result in an overestimation of the CFI, though the qualitative behavior remains intact (See SM~\cite{Supp}).
 
\begin{figure}[t]
\centering
\includegraphics[width=\linewidth]{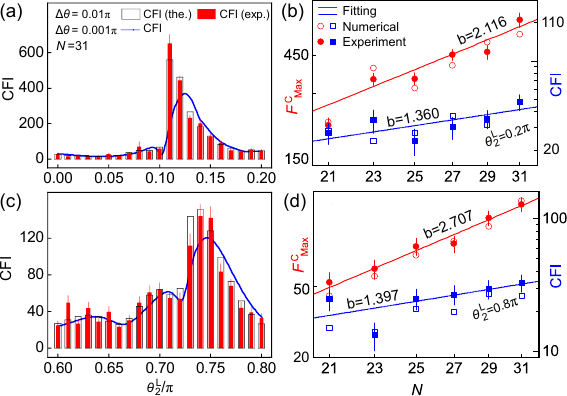}
\caption{\textbf{Experimental results for criticality-enhanced sensing}.
(a-b) Point gap closing case with $\theta_1^{L}{=}\theta_2^{R}{=}0.9 \pi$, $\theta_2^{L}{=}\theta_1^{R}$ and initial state $|\Psi_0\rangle{=}|0\rangle{\otimes}|V\rangle$. 
(a) CFI as a function of $\theta_2^{L}$. Maximum CFI $F^C_{\rm max}$ is observed near criticality, at $\theta_2^{L} {=} 0.11\pi$.
(b) Scaling of $F^C_{\rm max}$ with $N$ and CFI evaluated at $\theta_2^{L}{=}0.2\pi$, away from criticality.
(c-d) Line gap closing case with $\theta_1^{L}{=}\theta_2^{R}{=}0.05 \pi$, $\theta_2^{L}{=}\theta_1^{R}$ and $|\Psi_0\rangle{=}|0\rangle {\otimes} \left(|H\rangle {-} |V\rangle \right)/\sqrt{2}$. 
(c) CFI as a function of $\theta_2^{L}$. $F^C_{\rm max}$ is observed near criticality, at $\theta_2^{L}{=}0.74\pi$. 
(d) Scaling of $F^C_{\rm max}$ with $N$ and CFI at $\theta_2^{L}{=}0.8\pi$, away from criticality. 
All error bars indicate statistical uncertainties, estimated for Poissonian counting statistics (see SM~\cite{Supp}).}
\label{fig:exp}
\end{figure}

In Fig.~\ref{fig:exp}(a), we present the measured CFI of the evolved state $|\Psi_N\rangle$ for $N{=}31$ (requiring $T{=}15$ steps of the quantum walk), obtained for different values of $\theta_2^{L}$ near the point gap closing. 
The CFI peaks around $\theta_2^{L} {=} 0.11\pi$. 
These results are in good agreement with theoretical predictions and occur near the point gap closing phase transition at $\theta_2^{L} {=} 0.1 \pi$. 
As shown in Fig.~\ref{fig:exp}(b), the maximum CFI grows with $N$, which can be fitted as $F^C_{\rm max}{\sim} N^b$, with exponent $b{=}2.116$.  
For comparison, we also analyze the cases where $\theta_2^{L}$ deviates from the critical point. 
As shown in Fig.~\ref{fig:exp}(b), we set $\theta_2^L{=}0.2\pi$, fairly far from the critical point. 
The growth of the CFI with $N$ becomes significantly slower, with exponent $b{=}1.360$. 
The peak position and two distinct scaling behaviors clearly exhibit quantum-enhanced sensitivity and its demise away from criticality.

\begin{figure}[t]
\centering
\includegraphics[width=\linewidth]{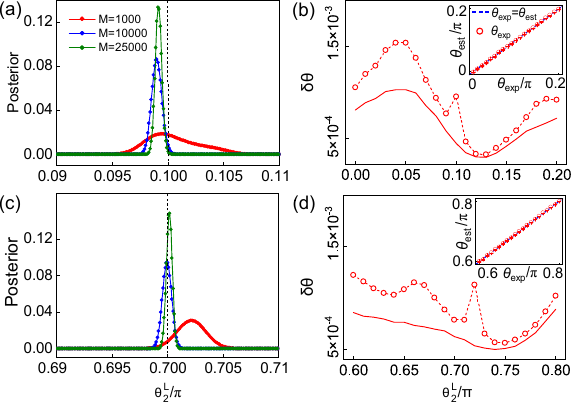} 
\caption{\textbf{Bayesian estimation from the experimental data}.  
(a-b) Point gap closing. 
(a) Posterior updates with measurement repetition numbers $M$. 
(b) Estimated value $\theta_{\rm est}$ for $\theta_2^L$ versus the experimentally set value $\theta_{\rm exp}$ with $M {=} 25000$. 
(inset) Precision of estimation in terms of standard deviation $\delta \theta$ and its lower bound given by the QFI (solid line). 
(c-d) Line gap closing. 
(c) Posterior updates with $M$. 
(d) Estimation $\theta_{\rm est}$ for $\theta_2^L$ versus $\theta_{\rm exp}$ with $M {=} 25000$.
(inset) Precision $\delta \theta$ and its lower bound given by the QFI (solid line).
Other relevant parameters are same as those in Fig.~\ref{fig:exp}.}
\label{fig:est}
\end{figure}

For line gap closing, Fig.~\ref{fig:exp}(c) presents the measured CFI near the criticality, with the maximum occurring around $\theta_2^{L} {=} 0.74\pi$.
This is adjacent to the theoretically predicted critical point at $\theta_2^{L} {=} 0.779\pi$. 
The maximum CFI again exhibits rapid growth with system size near the phase transition point. 
A fit to the experimental data yields a growth exponent of $b{=}2.707$, as shown in Fig.~\ref{fig:exp}(d). 
However, when $\theta_2^{L}$ deviates from the critical point, the growth of the CFI becomes significantly slower with $b{=}1.397$, as shown in Fig.~\ref{fig:exp}(d).
Additionally, our experimental system can also show another kind of enhanced sensitivity associated with a distinct type of energy gap closing, which is discussed in SM~\cite{Supp}.

% =============================================================================
\emph{Bayesian analysis.---} 
While Fisher information only provides a bound to the precision, we now employ Bayesian analysis to estimate the parameter and calculate the precision from the experimental data. 
The methodology is discussed in \textit{End Matter}.
The results are shown in Fig.~\ref{fig:est}, with (a-b) and (c-d) corresponding to point gap and line gap closing cases, respectively.
The improvement in both estimation and precision with increasing measurement number $M$ in the experiment is shown by the posteriors plotted in Figs.~\ref{fig:est}(a, c), when the real value is near the critical point, for the two different enhancement mechanisms.
The precision $\delta \theta$, represented by the standard deviation of the posterior distribution, reaches the minimum near the critical points, as shown in Figs.~\ref{fig:est}(b, d) for $M {=} 25000$. 
The computed precision values expectedly lie above the lower bound given by the QFI and further discrepancies can be attributed to small experimental imperfections and non-optimality of the measurement setup. 
As shown in the insets, the estimated value $\theta_{\rm est}$ faithfully follows the experimentally set value $\theta_{\rm exp}$.

% =============================================================================
\emph{Conclusion.---}
Experimental realization of criticality-enhanced sensitivity has remained a major challenge due to the difficulties of reaching the critical ground state or steady state. 
In this work, we experimentally demonstrate such sensitivity for estimating a bulk parameter in an NH topological system. 
Designing a NH sensor with quantum-enhanced precision is advantageous for estimating phase shifts induced by electric or magnetic fields and gradient fields such as gravity, as they naturally incorporate non-unitary evolution that occurs in open quantum systems.
Using a photonic quantum-walk platform, we observe quantum enhancement at two distinct topological phase transitions, accompanied by point and line gap closing. 
While the enhancement has been accomplished at the transient times, before reaching the steady state, our theoretical analysis confirms that it directly reflects the steady-state critical behavior. 
Furthermore, we validate the sensing capability of our platform through Bayesian analysis, which reveals that excellent estimation and precision can indeed be achieved. 
Fundamentally, our results reveal the role of different types of gap closings in achieving quantum enhanced sensitivity in NH systems, and experimentally demonstrate the resource efficiency of the probe through superlinear scaling with respect to system size. 
Practically, these results establish NH topological systems as a scalable platform for exploiting quantum criticality, and represent an important step towards realizing quantum-enhanced sensing in real-world applications.
To scale up our protocol to larger system sizes, one can leverage time-multiplexed architectures~\cite{Lin2022Observation} to reduce decoherence, and overcome the effect of photon loss by tuning the system to operate near a closed imaginary gap~\cite{xiao2024observation}, where the impact of loss is reduced.

% =============================================================================
\begin{acknowledgments}
Discussion with Tiangshu Deng is warmly acknowledged.
This work has been supported by the National Key R$\&$D Program of China (Grant No. 2023YFA1406701), the National Natural Science Foundation of China (Grants No.~W2433012, No.~W2541020, No.~12274059,  No.~12574528, No.~1251101297, No.~12025401, No.~92265209, No.~92476106, No.~12474352) and the Beijing National Laboratory for Condensed Matter Physics (Grant No.~2024BNLCMPKF010).
\end{acknowledgments}

% =============================================================================
\emph{Data availability.---}
The data that support the findings of this article are openly available on \href{https://github.com/SaubhikSarkar/QFI_QW}{Github}.

% =============================================================================
\section*{End Matter}

\emph{Details of the experimental implementation.---}
As shown in Fig.~\ref{fig:setup}, the single photons are generated by a pulsed laser with a central wavelength of 810~nm, a pulse duration of 88~ps, and a repetition rate of 20~MHz.
The pulses are attenuated to the single-photon level using NDFs, and the initial mean photon number per pulse is given by $\langle n\rangle {\approx 0.4}$.
Subsequently, PBS and HWP are used to prepare the initial coin state as $\ket{\psi_c}$. 
By setting the angle of HWP to $0$, $-22.5^\circ$ and $45^\circ$, the coin state $\ket{\psi_c}$ is prepared in $\ket{H}$, $(\ket{H}-\ket{V})/\sqrt{2}$, and $\ket{V}$, respectively.
The non-unitarity is introduced via controllable photon loss. 
This photon loss is realized using a PPBS, which reflects the $\ket{V}$ state with probability $p$ while fully transmitting the $\ket{H}$ state.
This process corresponds to an effective loss operator $\Gamma$, with loss strength $\gamma {=} -\frac{1}{4}\ln(1 - p)$.
The position-dependent coin operator $R_{\theta}$ is implemented by individually inserting two HWPs into the corresponding spatial modes.
To realize the spatially varying coin operators, the angles of the HWPs are set differently in two regions: in the left region ($j {\leq} {-}1$), the HWPs are fixed at $0$ and $\theta^{L}_{1,2}/4$, while in the right region ($j {\geq} 0$), the HWPs are fixed at $0$ and $\theta^{R}_{1,2}/4$, respectively.
This configuration effectively constructs a domain wall between the two regions, as illustrated in Fig.~\ref{fig:scheme}(c).
The shift operator $S$ is realized using a birefringent BD, which transmits vertically polarized photons without deviation, while horizontally polarized photons experience a lateral displacement of 3~mm into an adjacent spatial mode. 
After each step, photon counts at each position are recorded by APDs.
To ensure sufficient statistical significance, we collect a peak count of ${\sim} 10^4$ photons within 10 minutes, during which the interferometer visibility for each step is maintained above $99.5\%$, and the uncertainty of the wave plates is kept below $0.1^{\degree}$ (see SM~\cite{Supp}).

% =============================================================================
\emph{Bayesian estimation.---}
In our experiment, the position measurements, repeated $M$ times for a system size $N$, produces outcomes $\mathbf{m}{=}(m_{-n},\cdots,m_n)$, where the $j$-th outcome is obtained $m_j$ times, with $\sum_j m_j {=} M$.
The model probability of each outcome is given in terms of the state $\ket{\Psi_N(\theta_2^L)}$ and operator $\Pi_j$ as $p_j(\theta_2^L) {=} \braket {\Psi_N(\theta_2^L)|\Pi_j|\Psi_N(\theta_2^L)}$.
This is compared with the data $\mathbf{n}$ through the \textit{likelihood} function, given by the multinomial distribution function $P(\mathbf{n} | (\theta_2^L)) {=} \frac{M!}{\prod_j m_j !} \prod_j p_j^{m_j}$.
An initial knowledge of the parameter is given by the \textit{prior} $P(\theta_2^L)$.
If no prior information is known except for the ranges, i.e.,~$\theta_2^L {\in} \left[\theta_{\rm{min}},\theta_{\rm{max}} \right]$, then the prior distribution is flat $P(\theta_2^L) = \prod \frac{1}{\theta_{\rm{max}} - \theta_{\rm{min}}}$.
Now Bayes' theorem can be applied to write down the \textit{posterior}, which is the conditional probability distribution $P(\theta_2^L | \mathbf{n}) {=} P(\mathbf{n} | \theta_2^L) \, P(\theta_2^L)$, up to a normalization factor.
In the limit of large $M$, the central limit theorem applies, and the posterior takes a Gaussian form, the mean and variance of which give the estimated parameter values and precision, respectively.

%=============================================================================
\bibliography{Ref}

% =============================================================================
\beginsupplement

\begin{center}
\textbf{\large SUPPLEMENTAL MATERIAL:\\ Observation of Criticality-Enhanced Quantum Sensing in Nonunitary Quantum Walks}

\bigskip

Lei Xiao,$^{1,2}$ Saubhik Sarkar,$^{3,4}$ Kunkun Wang,$^{5}$ Abolfazl Bayat,$^{3,4,6}$ and Peng Xue$^{1}$ \\

\medskip

$^{1}${\small \em School of Physics, Southeast University, Nanjing 211189, China,} \\
$^{2}${\small \em Key Laboratory of Quantum Materials and Devices of Ministry of Education, Southeast University, Nanjing 211189, China} \\
$^{3}${\small \em Institute of Fundamental and Frontier Sciences, University of Electronic Science and Technology of China, Chengdu 610051, China} \\
$^{4}${\small \em Key Laboratory of Quantum Physics and Photonic Quantum Information, Ministry of Education,} \\
      {\small \em University of Electronic Science and Technology of China, Chengdu 611731, China} \\
$^{5}${\small \em School of Physics and Optoelectronics Engineering, Anhui University, Hefei 230601, China} \\
$^{6}${\small \em Shimmer Center, Tianfu Jiangxi Laboratory, Chengdu 641419, China} \\
\end{center}

% =============================================================================
\subsection{Parameter estimation}

In the single parameter estimation problem, an unknown parameter $\theta$ is first encoded in a probe's state $\rho_\theta$. 
The probe is then subjected to a measurement process, the outcomes of which generate a classical probability distribution.   
From this distribution, one can infer the unknown parameter with an estimator function.
The CFI $F^C$ can be used to write down the upper bound of the precision according to the Cram\'er-Rao inequality.
On the other hand, one can maximize $F^C$ with respect to all possible measurement bases to obtain the QFI $F^Q$, which is a measurement-independent quantity and quantifies the ultimate achievable precision limit. 
The QFI can be written as $F^Q {=} \text{Tr}\left[\mathcal{L}_{\theta}^2 \rho_\theta \right]$, where the symmetric logarithmic derivative operator $\mathcal{L}_\theta$ is defined implicitly as $\partial_{\theta}\rho_{\theta} {=} (\rho_\theta \mathcal{L}_\theta {+} \mathcal{L}_\theta \rho_\theta)/2$. 
For pure states $\rho_{\theta} {=} \ket{\psi_{\theta}} \bra{\psi_{\theta}}$, one gets $\mathcal{L}_{\theta} {=} 2 \partial_{\theta} \rho_{\theta}$ and QFI can be simplified to the expression mentioned in the main text.
The QFI corresponds to the CFI calculated with an optimal basis.
Although the optimal measurement basis is not unique, one choice is always given by the eigenbasis of $\mathcal{L}_\theta$. 
This sensing formalism with quantum systems can be adopted in NH systems too.
If the probe state is a non-unitarily evolved state or a right eigenstate of an NH Hamiltonian, it needs to be normalized by conventional norms so that the probability distributions obtained from the measurement procedures are normalized.
One can then define a valid density operator, and the standard procedure of defining the CFI and the QFI can be carried out with the same expressions.

% =============================================================================
\subsection{Generalized Brillouin zone}

The generalized Brillouin zone (GBZ) provides a theoretical formalism to understand the edge localization of the eigenstates due to skin effect in non-Hermitian systems in the presence of a point gap~\cite{Yao2018Edge, Yokomizo2019Non}.
In our work, the non-unitary operator $U$ and the underlying non-Hermitian Hamiltonian $H_{\rm eff}$ share the same eigenstates.
It is convenient to rewrite~\cite{Xiao2020Non}
\begin{align}
    U = \sum_{j} \left[ \ket{j-1} \bra{j} \otimes C_{-1}(j)  + \ket{j+1} \bra{j} \otimes C_{+1}(j) \right] ,
\end{align}
where $C_{\pm 1}(j)$ act on the coin state as
\begin{align}
    C_{-1}(j) &= R_c(\frac{\theta_2(j-1)}{2}) \; \ket{H}\bra{H} \; R_c(\frac{\theta_1(j)}{2}) \; \Gamma_c , \nn
    C_{+1}(j) &= R_c(\frac{\theta_2(j+1)}{2}) \; \ket{V}\bra{V} \; R_c(\frac{\theta_1(j+1)}{2}) \; \Gamma_c ,
\end{align}
with $R_c(\theta) = e^{-i \theta \sigma_y}$ and $\Gamma_c =  e^{\gamma \sigma_z}$.
In our domain wall system, $\theta_{1,2}(j) = \theta_{1,2}^L$ for $-n {\le} j {\le} -1$ and $\theta_{1,2}(j) = \theta_{1,2}^R$ for $0 {\le} j {\le} n$.
Now, following the GBZ formalism, the ansatz for an eigenstate of $U$ in the domain wall system is written as~\cite{deng2019non}
\begin{align}
    \ket{\psi} &= \sum_{j=-n}^{-1} \beta_L^j \left[\chi_{H,L} \ket{j, H} + \chi_{V,L} \ket{j, V} \right]
                 + \sum_{j=0}^{n} \beta_R^j \left[\chi_{H,R} \ket{j, H} + \chi_{V,R} \ket{j, V} \right] \nn
               &\equiv \sum_{j=-n}^{-1} \beta_L^j \ket{j} \otimes \ket{\chi_{L}} + \sum_{j=0}^{n} \beta_R^j \ket{j} \otimes \ket{\chi_{R}},
\end{align}
with $\beta_{\alpha}$ denoting the localization in the left and right parts for $\alpha = L, R$, respectively, and $\ket{\chi_{\alpha}}$ being the coin state.
Now the eigen-equation $U \ket{\psi} = \lambda \ket{\psi}$ in the bulk of each part leads to
\begin{align}
    \begin{pmatrix}C_{-1, \alpha} \beta_{\alpha} + \frac{C_{+1, \alpha}}{\beta_{\alpha}} \end{pmatrix} \ket{\chi_{\alpha}}
    = \lambda \ket{\chi_{\alpha}} ,
    \label{eq:chi}
\end{align}
where $C_{\pm 1, \alpha}$ acts in the bulk of $\alpha$ part of the system, i.e., $C_{\pm 1, L} = C_{\pm 1}(j)$ with $j {\in} [-n+1, -2]$ and $C_{\pm 1, R} = C_{\pm 1}(j)$ with $j {\in} [1, n-1]$.
The condition for non-trivial solutions is
\begin{align}
    \det \begin{pmatrix} C_{-1, \alpha} \beta_{\alpha} + \frac{C_{+1, \alpha}}{\beta_{\alpha}} - \lambda \mathds{1} \end{pmatrix} = 0,
    \label{eq:beta}
\end{align}
which is a quadratic equation for $\beta_{\alpha}$, as $C_{\pm 1, \alpha}$ have zero determinants.
The solutions $\beta_{1, \alpha}$ $\beta_{2, \alpha}$ (for a given value of $\lambda$) can be used to write down the eigenstate  as a superposition
\begin{align}
    \ket{\psi} &= \sum_{m=1}^2 c_m \; \Bigg(\sum_{j=-n}^{-1} \beta_{m,L}^j \ket{j} \otimes \ket{\chi_{L}} + \sum_{j=0}^{n} \beta_{m,R}^j \ket{j} \otimes \ket{\chi_{R}} \Bigg) \nn
               &= \sum_{m=1}^2 \Bigg(\sum_{j=-n}^{-1} \beta_{m,L}^j \ket{j} \otimes \ket{\phi_{m, L}} + \sum_{j=0}^{n} \beta_{m,R}^j \ket{j} \otimes \ket{\phi_{m, L}} \Bigg),
    \label{eq:state}
\end{align}
where $\ket{\phi_{m, \alpha}} = c_m \ket{\chi_{\alpha}}$.
Now we apply Eq.~\eqref{eq:state} at the edges between sites $j=-n, n$ and sites $j=0, 1$ and use Eq.~\eqref{eq:chi} to simplify the results.
Specifically, at $j=-n$, Eq.~\eqref{eq:state} leads to
\begin{align}
    \ket{-n} \otimes \Bigg[C_{-1,L} \left(\beta_{1,L}^{-n+1} \ket{\phi_{1,L}} + \beta_{2,L}^{-n+1} \ket{\phi_{2,L}} \right)
                           C_{+1}(-n) \left(\beta_{1,L}^{n} \ket{\phi_{1,R}} + \beta_{2,R}^{n} \ket{\phi_{2,L}} \right) \Bigg]
    = \ket{-n} \otimes \lambda \left[\beta_{1,L}^{-n+1} \ket{\phi_{1,L}} + \beta_{2,L}^{-n+1} \ket{\phi_{2,L}} \right] \nn
    \implies - C_{+1,L} \beta_{1,L}^{-n-1} \ket{\phi_{1,L}} - C_{+1,L} \beta_{2,L}^{-n-1} \ket{\phi_{2,L}}
             + C_{+1}(-n) \beta_{1,R}^{n} \ket{\phi_{1,R}} + C_{+1}(-n) \beta_{2, R}^{n} \ket{\phi_{2,R}} = 0 \qquad (\text{using Eq.~\eqref{eq:chi}})
\end{align}

\begin{figure}[t]
\centering
\includegraphics[width=0.5\linewidth]{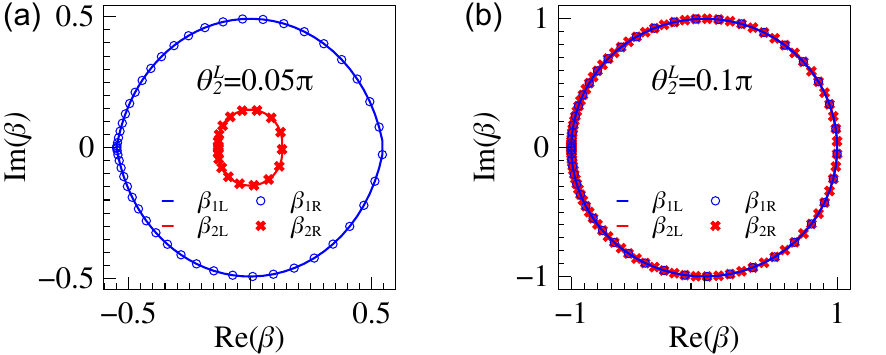}
\caption{\textbf{GBZ and point gap}. The fixed parameters are $\gamma {=} 0.3$ and $\theta_1^L {=} \theta_2^R {=} 0.9 \pi$.
(a) Point gap at $\theta_2^L {=} 0.05 \pi$ results in GBZ with $|\beta{m, \alpha}| {\ne} 1$. (b) Point gap closing at $\theta_2^L {=} 0.1 \pi$ results in GBZ with $|\beta{m, \alpha}| {=} 1$. Here $m {=} 1, 2$ for the two solutions in the left and right region denoted by $\alpha {=} L, R$, respectively.}
\label{fig:GBZ}
\end{figure}

Applying this strategy for the other three edges, we derive a set of linear coupled equations written as $\mathcal{M} \begin{pmatrix} \ket{\phi_{1,L}} & \ket{\phi_{2,L}} & \ket{\phi_{1,R}} & \ket{\phi_{2,R}} \end{pmatrix}^T = 0$, with
\begin{align}
    \mathcal{M} =
    \begingroup
    \setlength\arraycolsep{2pt}
    \begin{pmatrix}
       - \frac{C_{+1,L}}{\beta_{1,L}^{n+1}} & - \frac{C_{+1,L}}{\beta_{2,L}^{-n-1}} & C_{+1}(-n) \beta_{1,R}^{n} & C_{+1}(-n) \beta_{2, R}^{n} \\
       \frac{C_{+1,L}}{\beta_{1,L}^{2}} - \frac{\lambda}{\beta_{1,L}} & - \frac{C_{+1,L}}{\beta_{2,L}^{2}} - \frac{\lambda}{\beta_{2,L}}  & C_{-1}(-1) & C_{-1}(-1)  \\
       \frac{C_{+1}(0)}{\beta_{1,L}} & \frac{C_{+1}(0)}{\beta_{2,L}} & -\frac{C_{+1,R}}{\beta_{1,R}} & \frac{C_{+1,R}}{\beta_{2,R}}\\
       \frac{C_{-1}(n)}{\beta_{1,L}^{n}} & \frac{C_{-1}(n)}{\beta_{2,L}^{n}} & C_{+1,R} \beta_{1,R}^{n{-}1} {-} \lambda \beta_{1,L}^n & C_{+1,R} \beta_{2,R}^{n{-}1} {-} \lambda \beta_{2,L}^n
    \end{pmatrix} .
    \endgroup
\end{align}
Again, the non-triviality of the solutions demands $\det \mathcal{M} {=} 0$.
One can now follow the procedure developed in Refs.~\cite{deng2019non, Xiao2020Non} to write down the equation for the GBZ in the asymptotic limit $n \to \infty$ as
\begin{align}
    \zeta (\beta_{1,L},\beta_{2,L},\beta_{1,R},\beta_{2,R}) = 0 ,
    \label{eq:zeta}
\end{align}
with the definition
\begin{equation}
    \zeta(\beta_{1,L},\beta_{2,L},\beta_{1,R},\beta_{2,R}):=
    \begin{cases}
    |\beta_{1,L}\beta_{1,R}|-1, & |\beta_{2,L}\beta_{2,R}|\leqslant1\,\,\text{and}\,\,|\beta_{2,L}\beta_{1,R}|\leqslant1\,\,\text{and}\,\,|\beta_{1,L}\beta_{2,R}|\leqslant1,\\
    |\beta_{2,L}\beta_{2,R}|-1, & |\beta_{1,L}\beta_{1,R}|\geqslant1\,\,\text{and}\,\,|\beta_{1,L}\beta_{2,R}|\geqslant1\,\,\text{and}\,\,|\beta_{2,L}\beta_{1,R}|\geqslant1,\\
    |\beta_{1,L}|-|\beta_{2,L}|, & |\beta_{2,L}\beta_{1,R}|\geqslant1\,\,\text{and}\,\,|\beta_{1,L}\beta_{2,R}|\leqslant1,\\
    |\beta_{1,R}|-|\beta_{2,R}|, & |\beta_{1,L}\beta_{2,R}|\geqslant1\,\,\text{and}\,\,|\beta_{2,L}\beta_{1,R}|\leqslant1 .
    \end{cases}
\label{eq:zeta_def}
\end{equation}
One can express $\beta{m, \alpha}$ in terms of $\lambda$ and use these equations for $\zeta$ to solve for the eigenenergies.
Then $\beta{m, \alpha}$ can be derived using Eq.~\eqref{eq:chi}, the trajectories of which generates the GBZ in the complex plane.
We show this in Fig.~\ref{fig:GBZ}(a) where $|\beta{m, \alpha}| {\ne} 1$ results in the edge-localization of eigenstates (i.e., the skin effect).
%The results match well with the GBZ computed with numerically evaluated eigenenergies in finite systems.
In Fig.~\ref{fig:GBZ}(b), we see that the GBZ is represented by a unit circle as we choose point gap closing parameter values.
Here $|\beta{m, \alpha}| {=} 1$ signals delocalization of eigenstates which suppresses the skin effect.
This provides analytical support to the corresponding criticality condition of $\theta_{1}^{L} {+} \theta_{2}^{L} {=} \pi$, reported in the main text.

% =============================================================================
\subsection{Transient time scaling}

\begin{figure}[t]
\centering
\includegraphics[width=0.75\linewidth]{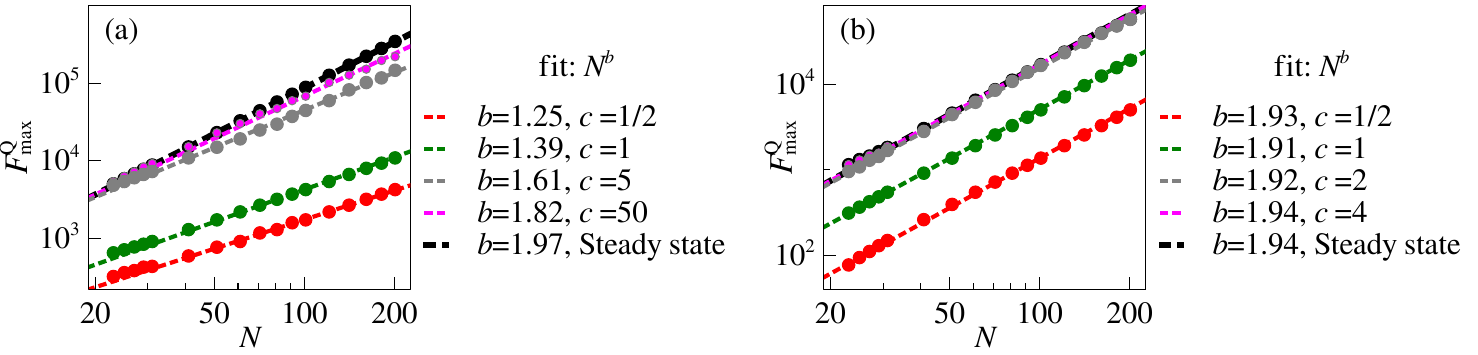}
\caption{\textbf{Scaling at different timescales}. (a) Point gap closing case with $\gamma {=} 0.3$ and $\theta_1^L {=} \theta_2^R {=} 0.9 \pi$. (b) Line gap closing case with $\gamma {=} 0.3$ and $\theta_1^L {=} \theta_2^R {=} 0.05 \pi$.}
\label{fig:time}
\end{figure}

The favorable scaling of QFI with system size near the critical point of the non-Hermitian systems should be appearing for the steady state.
Therefore, in principle, one needs to investigate the scaling properties in the asymptotic time limit.
However, in practice, we are often limited to finite evolution time available in the experiments.
One important feature of the model simulated in our experiment is that the quantum-enhanced scaling can be observed even before reaching the asymptotic limit. 
In the main text, we show how the steady state value of the QFI is reached in Figs.~2(b) and 2(f) near the closure of point gap and the line gap, respectively.
Here we provide the scaling behavior at different timescales and show the scaling exponents at those times.
We consider the evolution times to be $T = c (N-1)$, whereas our experiment corresponds to $c = 1/2$.
For the point gap closing case, shown in Fig.~\ref{fig:time}(a), we observe the scaling exponent gradually approaching the quadratic behavior as $c$ is increased.
For the line gap closing case, the steady state behavior is reached for much smaller values of $c$.
As shown in Fig.~\ref{fig:time}(b), the scaling exponents reflect close to quadratic scaling even for small $c$ values.

% =============================================================================
\subsection{Error analysis}

In addition to the statistical fluctuations inherent to Poissonian photon counts, several experimental imperfections may influence the results. 
These include photon scattering and reflection at optical interfaces, inaccuracies in the setting angles of wave plates, and spatial misalignment between paired beam displacers that form interferometric paths, which can introduce decoherence. 
These imperfections eventually accumulate and manifest in the probability distribution of the quantum walk.

The error associated with the CFI $F^C(\theta_i)$ is derived using standard error propagation,
\begin{align}
\Delta F^C(\theta_i) = \sqrt{\sum_j \left[ \left( \frac{\partial F^C(\theta_i)}{\partial \tilde{p}_j(\theta_{i+1})} \Delta P_{i+1}(j) \right)^2 + \left( \frac{\partial F(\theta_i)}{\partial \tilde{p}_j(\theta_{i})} \Delta \tilde{p}_j(\theta_{i}) \right)^2 \right]},
\end{align}
where $\Delta \tilde{p}_j = \sqrt{ \sum \left[ \left( \frac{\partial \tilde{p}_j}{\partial m_j} \right)^2 \Delta m_j^2 \right] }$, and $\Delta m_j$ is the statistical uncertainty associated with single-photon counting.

\begin{figure}[t]
\centering
\includegraphics[width=0.7\linewidth]{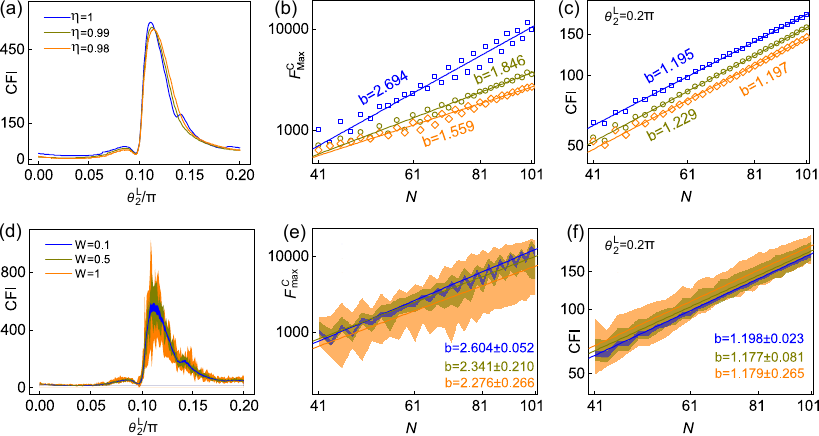}
\caption{\textbf{Simulated Fisher information under experimental imperfections.} The point gap closing with $\theta_{1}^L=\theta_{2}^R=0.9\pi$, $\theta_1^R=\theta_2^L$, initial state $|\Psi_0\rangle{=}|0\rangle{\otimes}|V\rangle$ and $\gamma=0.3$. 
(a) CFI as a function of $\theta_2^{L}$ with different $\eta$.
(b)-(c) Scaling of $F^C_{\rm max}$ with different $\eta$ and CFI evaluated at $\theta_2^{L}{=}0.2\pi$, away from criticality.
(d) CFI as a function of $\theta_2^{L}$ with different $W$.
(e)-(f) Scaling of $F^C_{\rm max}$ with different $W$ and CFI evaluated at $\theta_2^{L}{=}0.2\pi$, away from criticality. The shaded region represents the standard deviations averaged over $20$ random simulations}.
\label{Fig:align}
\end{figure}

In our experiment, several practical imperfections may influence the results. These include the scattering and reflection of photons by optical elements,as well as the inaccurate photon loss introduced by PPBS, the spatial misalignment of BDs within the interferometric setup, which can introduce decoherence, and the inaccuracy of wave plate. To quantify their impact, we numerically simulated the effect of each imperfection individually.
First, with respect to photon loss, the measured photon counts are normalized to obtain Fisher information. The absolute photon loss does not affect the scaling behavior in our work. Misalignment between pairs of BDs in the interferometer can cause decoherence. For simplicity, we model this as a depolarizing channel per step, described by $\rho'(T)=\eta\rho(T)+(1-\eta)(I \otimes\sigma_z) \rho(T) (I\otimes\sigma_z)^\dagger$, where $\rho(T)=\ket{\Psi_T} \bra{\Psi_T}$, $\eta$ parameterizes the decoherence strength. As illustrated in Fig.~\ref{Fig:align} (a)-(c), under the influence of decoherence, the Fisher information exhibits noticeable shifts. In particular, the scaling behavior of $F_\text{max}$ shows degradation as the decoherence intensifies when $\eta<0.98$. For $\theta_2^{L}=0.2\pi$, decoherence predominantly induces a reduction in Fisher information magnitude, while preserving scaling characteristics. In our experiment, the typical $\eta\geq0.995$ makes decoherence irrelevant to conclusions.

The inaccuracy of the wave plates, which means that each wave plate can deviate from its ideal value, is indicated as $\theta'\in[\theta+W, \theta-W]$, where $W$ represents the extent of deviation. As shown in Fig.~\ref{Fig:align} (d)-(f), we performed 20 numerical simulations for different $W$ values, with the shaded regions representing the standard deviation of the Fisher information obtained. In our experiments, the typical $W$ remains below $0.1^{\degree}$, thus not affecting the conclusions.

Although these imperfections mentioned above cannot be entirely eliminated in practice, they only lead to minor deviations between the experimental data and theoretical predictions, without obscuring the key observations of criticality-enhanced quantum sensing.

\begin{figure}[b]
\centering
\includegraphics[width=0.8\linewidth]{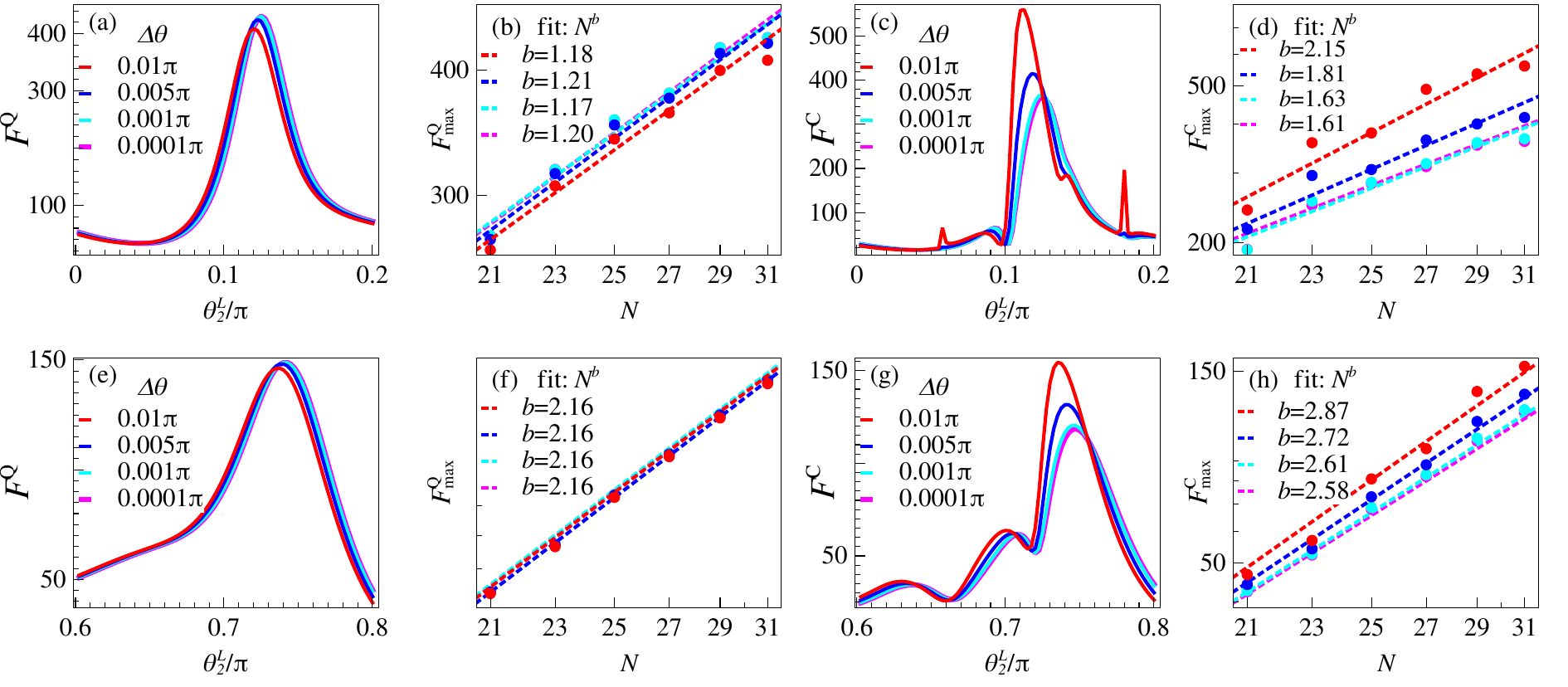}
\caption{\textbf{Convergence of QFI and CFI}. (top row) Point gap closing case with $\gamma {=} 0.3$ and $\theta_1^L {=} \theta_2^R {=} 0.9 \pi$. 
(bottom row) Line gap closing case with $\gamma {=} 0.3$ and $\theta_1^L {=} \theta_2^R {=} 0.05 \pi$.
(a, e) QFI as a function of $\theta_2^L$ for various $\Delta \theta$ values.
(b, f) Maximum QFI as a function of $N$.
(c, g) CFI as a function of $\theta_2^L$ for various $\Delta \theta$ values.
(d, h) Maximum CFI as a function of $N$.}
\label{fig:diff}
\end{figure}

One practical effect of the photon loss is to limit the number of steps achievable in the experiment. 
The number of injected photons is almost an abundant source as, we can pump up to $\sim 10^6$ photons per second in our experiment, and the entire setup remains stable over a detection window of about 20 minutes. 
Therefore, a sufficiently large number of photons can be injected into the interferometer for the measurement.
With a given number of injected  photons $N_{\rm ph}^{(0)}$, the number of photons after $T$ steps decay to $N_{\rm ph}^{(T)}\sim N_{\rm ph}^{(0)}\times e^{2(\text{Im}(E)_\text{Max}-\gamma)T}$, where $\text{Im}(E)_\text{Max}$ denotes the maximum imaginary part of the spectrum (excluding the boundary modes). 
Therefore, the procedure can reach $T=1/2(\text{Im}(E)_\text{Max}-\gamma)\times \log{N_{\rm ph}^{(0)}}$ steps while still having one photon left.

Another source of imperfection arises from finite resolution of rotational angles that can be implemented in the experiment.
This can lead to lack of convergence in determining the derivatives appearing in the expressions for Fisher information.
Therefore we study this issue and present the results in Fig.~\ref{fig:diff}, where the the top and bottom rows correspond to the point gap and line gap closing cases, respectively.
In the experiment, the smallest step size achievable is $\Delta \theta = 0.01 \pi$.
As shown in Figs.~\ref{fig:diff}(a) and (e), this step size is small enough to produce QFI quite close to the converged values.
Correspondingly, as Figs.~\ref{fig:diff}(b) and (f) shows, the scaling exponents for the peak QFI values are close to each other regardless of step size for the system sizes considered in the experiment.
The calculation of CFI with different step sizes reveal that although the resolution in the experiment is good enough to capture the peak at criticality and values close to converged ones away from criticality, it overestimates the CFI near the critical point (see Figs.~\ref{fig:diff}(c) and (g)).
This results in overestimation of the scaling exponents extracted from the system sizes used in the experiment, as shown in 
Figs.~\ref{fig:diff}(d) and (h).
Therefore, while the limitation in step size does not qualitatively affect the capture of critical enhancement, it results in scaling exponents to be quantitatively different.

% =============================================================================
\subsection{Additional mechanism for quantum-enhanced sensitivity}

\begin{figure}[t]
\centering
\includegraphics[width=0.5\linewidth]{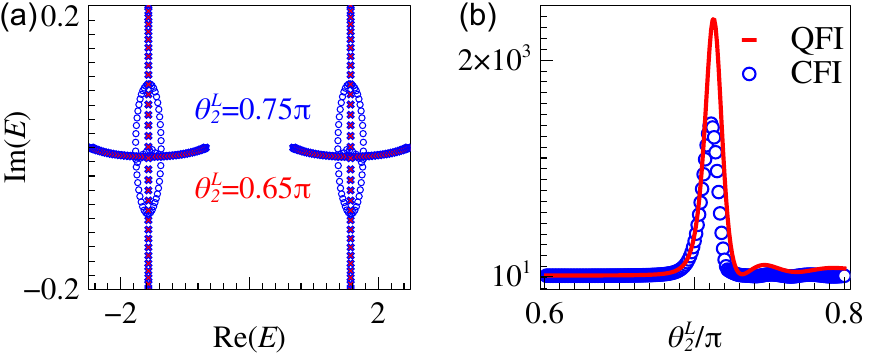}
\caption{\textbf{Additional mechanism for enhanced sensitivity}. The parameters are $\theta_1^{L}=0.45\pi$, $\theta_1^{R}=0.1 \pi$, $\theta_2^{R}=1.45\pi$, $\gamma = 0.3$.(a) Closing of point gap like structure in spectrum with OBC. (b) The QFI and CFI near the gap closing.}.
\label{fig:obc1}
\end{figure}

The criticality based quantum sensing described in the main text is inherently connected to the gap closing mechanism of the non-Hermitian system.
The eigenstates of a finite system with open boundary conditions (OBC) go through phase transitions corresponding to gap closing for closed boundary conditions.
The topological transitions are manifested through localization-delocalization transitions of the eigenstates.
However, in our quantum walk setup, it was found that a point gap like structure can appear in the complex spectrum under OBC as well.
Closing of such gap also led to enhanced sensitivity, although no definitive association to a localization-delocalization transition was found.
We show this in Fig.~\ref{fig:obc1} with the parameters $\theta_1^L {=} 0.45\pi, \theta_1^R {=} 0.1\pi, \theta_2^R {=} 1.45\pi$ and $\gamma {=} 0.3$ while $\theta_2^L$ is the parameter to be estimated.
Note that here we no longer have the constraint $\theta_{1}^{L} {=} \theta_{2}^{R}, \theta_{2}^{L} {=} \theta_{1}^{R}$.
In this case, we notice a point gap closing near $\theta_2^L {=} 0.65\pi$, as shown in Fig.~\ref{fig:obc1}(a).
We then proceed to calculate the QFI and CFI (with position measurements) for $\ket{\Psi_N}$ obtained after $T {=} (N-1)/2$ time steps to find characteristic peak in the vicinity, as shown in Fig.~\ref{fig:obc1}(b).

\begin{figure*}[t]
\centering
\includegraphics[width=\textwidth]{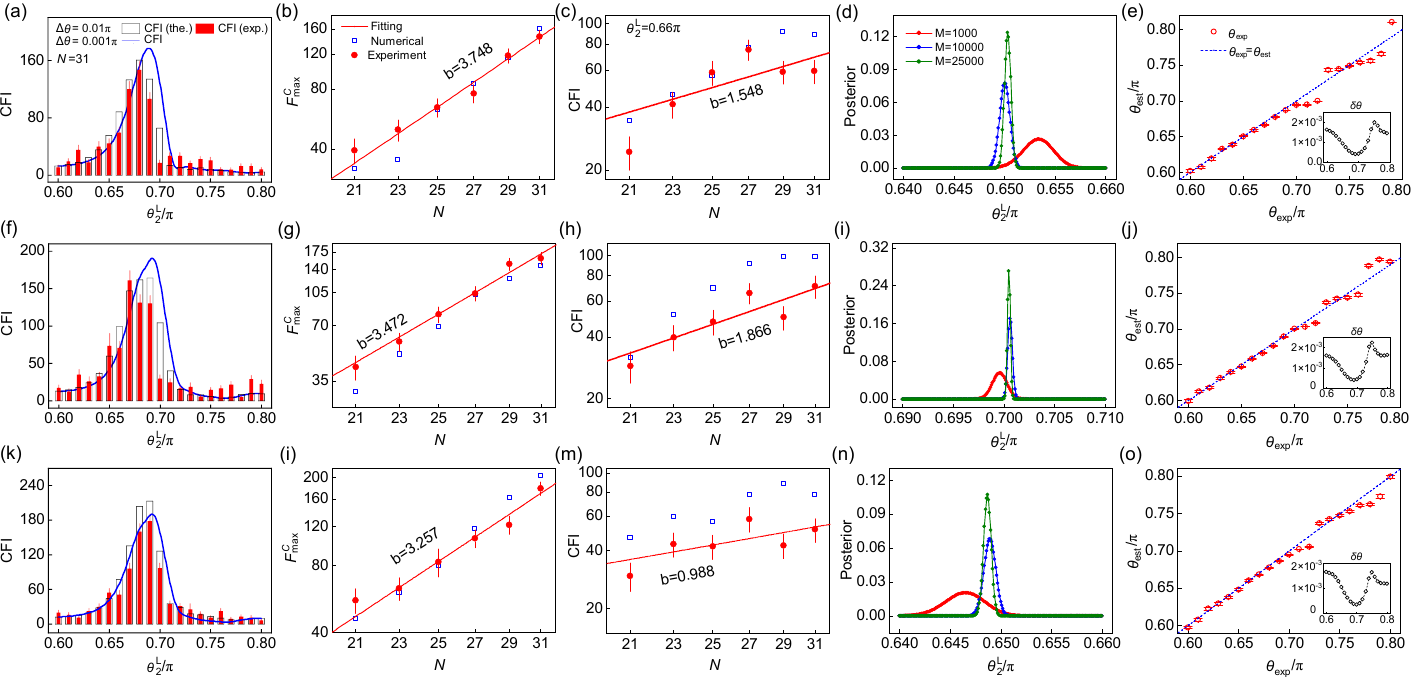}
\caption{\textbf{Experimental results for different initial states}. The point gap case under OBC with parameters $\theta_1^{L}{=}0.45\pi$, $\theta_1^{R}{=}0.1 \pi$, $\theta_2^{R}{=}1.45\pi$.
(a-e) The initial state is $|\Psi_0\rangle{=}|0\rangle\otimes |H\rangle$. (a) CFI as a function of $\theta_2^{L}$, peaking at $\theta_2^{L} {=} 0.68\pi$. (b) Scaling of the maximum CFI with system size $N$ as $F^C_{\rm max} {\sim} N^b$, with a fitted exponent $b{=}3.748$. (c) CFI at $\theta_2^{L}{=}0.66\pi$, with a fitted exponent $b{=}1.548$. (d) Posterior updates with measurement repetition numbers $M$. (e) Estimated value $\theta_{\rm est}$ for $\theta_2^L$ versus the experimentally set value $\theta_{\rm exp}$ with $M {=} 25000$.
(f-j) The initial state is $|\Psi_0\rangle{=}|0\rangle\otimes |V\rangle$. (f) CFI as a function of $\theta_2^{L}$, peaking at $\theta_2^{L} {=} 0.67\pi$. (g) Fitted exponent $b{=}3.472$. (h) CFI at $\theta_2^{L}{=}0.66\pi$, with a fitted exponent $b{=}1.866$. (i) Posterior updates with $M$. (j) Estimated value $\theta_{\rm est}$ for $\theta_2^L$ versus $\theta_{\rm exp}$ with $M {=} 25000$.
(k-o) The initial state is $|\Psi_0\rangle{=}|0\rangle\otimes |H\rangle+|V\rangle/\sqrt{2}$. (k) Maximum QFI as a function of $\theta_2^{L}$, peaking at $\theta_2^{L} {=} 0.69\pi$. (l) Scaling of the maximum CFI with system size $N$, with a fitted exponent $b{=}3.257$. (m) CFI at $\theta_2^{L}{=}0.66\pi$, with a fitted exponent $b{=}0.988$. (n) Posterior updates with $M$. (o) Estimated value $\theta_{\rm est}$ for $\theta_2^L$ versus $\theta_{\rm exp}$ with $M {=} 25000$.}
\label{fig:obc2}
\end{figure*}

Our experimental setup enables us to demonstrate this enhancement.
We further experimentally investigate the impact of different initial coin states on criticality-enhanced sensing, as illustrated in Fig.~\ref{fig:obc2}.
The initial coin states for Fig.~\ref{fig:obc2}(a-e), Fig.~\ref{fig:obc2}(f-j), and Fig.~\ref{fig:obc2}(k-o) are $|\Psi_0\rangle=|0\rangle\otimes |H\rangle$, $|\Psi_0\rangle=|0\rangle\otimes |V\rangle$, and $|\Psi_0\rangle=|0\rangle\otimes |H\rangle+|V\rangle/\sqrt{2}$ respectively.
We notice that changing the initial state affects the actual value of CFI obtained but has little effect on the peak position, scaling behaviour, and overall estimation and precision for large number of measurement repetitions.

% =============================================================================
\end{document}